\documentclass[10pt,twocolumn,letterpaper]{article}

\usepackage[pagenumbers]{cvpr} 

\usepackage{graphicx}
\usepackage{amsmath}
\usepackage{amssymb}
\usepackage{booktabs}
\usepackage{mathtools}
\usepackage[subtle]{savetrees}

\RequirePackage{algorithm}
\RequirePackage{algorithmic}

\DeclarePairedDelimiter\ceil{\lceil}{\rceil}
\DeclarePairedDelimiter\floor{\lfloor}{\rfloor}

\graphicspath{ {./Figures/} }

\usepackage[accsupp]{axessibility}
\usepackage[pagebackref,breaklinks,colorlinks]{hyperref}

\usepackage[capitalize]{cleveref}
\crefname{section}{Sec.}{Secs.}
\Crefname{section}{Section}{Sections}
\Crefname{table}{Table}{Tables}
\crefname{table}{Tab.}{Tabs.}

\def\cvprPaperID{8393} 
\def\confName{CVPR}
\def\confYear{2022}

\begin{document}

\title{Split Hierarchical Variational Compression}

\author{Tom Ryder\thanks{co-first author. The work of Tom Ryder is conducted during his employment at Huawei Technologies R\&D UK.} \hspace{0.5em} Chen Zhang$^{* \dagger}$ \hspace{0.5em} Ning Kang$^{\dagger}$ \hspace{0.5em} Shifeng Zhang$^{\dagger}$\\
$^{\dagger}$Huawei Noah's Ark Lab\\
{\tt\small \{chenzhang10, kang.ning2, zhangshifeng4\}@huawei.com}
}
\maketitle

\begin{abstract}
 Variational autoencoders (VAEs) have witnessed great success in performing the compression of image datasets. This success, made possible by the bits-back coding framework, has produced competitive compression performance across many benchmarks. However, despite this, VAE architectures are currently limited by a combination of coding practicalities and compression ratios. That is, not only do state-of-the-art methods, such as normalizing flows, often demonstrate out-performance, but the initial bits required in coding makes single and parallel image compression challenging. To remedy this, we introduce Split Hierarchical Variational Compression (SHVC). SHVC introduces two novelties. Firstly, we propose an efficient autoregressive prior, the autoregressive sub-pixel convolution, that allows a generalisation between per-pixel autoregressions and fully factorised probability models. Secondly, we define our coding framework, the autoregressive initial bits, that flexibly supports parallel coding and avoids -- for the first time -- many of the practicalities commonly associated with bits-back coding. In our experiments, we demonstrate SHVC is able to achieve  state-of-the-art compression performance across full-resolution lossless image compression tasks, with up to 100x fewer model parameters than competing VAE approaches. 
\end{abstract}

\section{Introduction}
\label{sec:intro}

The volume of data, measured in terms of IP traffic, is currently witnessing an exponential year-on-year growth ~\cite{forecast2019cisco}. Fuelled by the demand for high-resolution media content, it is estimated that 80\% of this data is in the form of images and video ~\cite{forecast2019cisco}. Data service providers, such as cloud and streaming platforms, have consequently seen costs associated with transmission and storage become prohibitively expensive. For example, an increased demand for streaming services forced major providers to throttle the maximum resolution of video content to 720p during the coronavirus pandemic. 
As such, these challenges have renewed the need for the development of high-performance data compression codecs.

One solution to this problem has been the development of approaches using likelihood-based generative models capable of discrete density estimation ~\cite{mentzer2019practical,townsend2019practical,hoogeboom2019integer,berg2020idf++,ho2019compression,townsend2019hilloc,kingma2019bit,mentzer2020learning,cao2020lossless,zhang2021ivpf, zhang2021iflow}. 
Such methods operate by learning a deep probabilistic model of the data distribution, which, in combination with entropy coders, can be used to compress data. Here, according to Shannon's source coding theorem \cite{mackay2003information}, the minimal required average codelength is bounded by the expected negative log-likelihood of the data distribution.

From this family of generative models, there have emerged three dominant modes for data compression: \textit{normalizing flows} \cite{hoogeboom2019integer,berg2020idf++, zhang2021ivpf, zhang2021iflow}, \textit{variational autoencoders} \cite{townsend2019hilloc,kingma2019bit,mentzer2020learning} and \textit{autoregressive models} ~\cite{salimans2017pixelcnn++,van2016conditional,jun2020distribution} \footnote{Most recently, score-based generative models have been adapted for data compression \cite{kingma2021variational}, but current approaches require an impractical number of operations at inference time.}. In fact, each of these approaches can be thought of as a traversal on the Pareto frontier of inference speed and compression performance. With broad generality, autoregressive models 
can often be the most powerful but the slowest; variational autoencoders are often the weakest but the fastest; and normalizing flows -- depending on the variant -- sit somewhere in between. 

In this paper, we consider data compression with VAEs, and focus on extending the efficient frontier; obtaining solutions faster than popular VAEs that achieve state-of-the-art compression ratios.
Use of VAEs, however, poses two outstanding challenges. Firstly,  we should achieve competitive coding ratios without greatly sacrificing time complexity. For example, best iterates currently require one of two ingredients to improve performance: building either a deep hierarchy of latent variables \cite{child2020very} or use of autoregressive priors \cite{pmlr-v70-reed17a, gulrajani2016pixelvae}.  The latter idea, especially popular in the codecs of the \textit{lossy} compression community \cite{NEURIPS2018_53edebc5}, posits a model that flexibly learns both local (via autoregression) and global (via hierarchical latent representation) data modalities (e.g. low-frequency information). Whilst these approaches, such as MS-PixelCNN \cite{pmlr-v70-reed17a} and PixelVAE \cite{gulrajani2016pixelvae}, have had some success in achieving more efficient trade-offs, generation of even moderately sized images is still to the order of minutes \cite{mentzer2019practical}.

Secondly, there should exist a practical means by which to efficiently perform single-image compression. Single-image compression then permits parallel coding, which is highly desirable. However, translating a VAE into a lossless codec is currently achieved using the \textit{bits-back} coding framework (predominantly, bits-back ANS), which requires a large number of \textit{initial bits} \cite{townsend2019hilloc, wallace1990classification, hinton93keeping} (see Section \ref{sec:bbans}). Whilst this is a trivial number of bits on large image datasets (where we can amortize this cost), it renders bits-back an impractical approach for single-image compression.
Furthermore, even large datasets are often coded such that images are interlinked. Access to a single image in the middle of a sequence would therefore require all prior images in the bitstream to be additionally decompressed. 

To that end, we propose two novelties for use in VAE-based compression designed to address these challenges. The first, our \textit{autoregressive sub-pixel convolution}, introduces a simple autoregressive factorisation -- not dissimilar from the transformations used in normalizing flows \cite{berg2020idf++, zhang2021ivpf, zhang2021iflow} -- designed to present an efficient interpolation between fully-factorised probability distributions and the impractical per-pixel autoregressions. Built from a modified space-to-depth convolution operator, we losslessly downsample data variables before performing a computationally efficient autoregression along the channel dimension. Our autoregressive operator is then advantaged by a number of network evaluations invariant to data dimensions, with each autoregression crucially performed on a downsampled version on the input tensor. More broadly, we view this framework as a generalisation of many popular autoregressive ``context" models used in data compression \cite{salimans2017pixelcnn++, DBLP:conf/icip/MinnenS20, gulrajani2016pixelvae, Zhang_2020_ACCV}.

Our second contribution, \textit{autoregressive initial bits}, presents a general framework for avoiding the impracticalities of bits-back ANS, allowing for eminently parallelizable coding. This technique, highly compatible with our autoregressive model, partitions the data variable into two \textit{splits} such that the second partition is conditionally independent of the latent variable(s), given the first. In this way, we illustrate how we can use the entropy coding of the conditionally independent partition to both \textit{supply} and \textit{remove} the initial bits necessitated by bits-back ANS. We demonstrate that this approach reduces the bit overhead on a per-image basis by close to 20x. 

Finally, we combine the above contributions to present our codec, \textit{Split Hierarchical Variational Compression} (SHVC). SHVC posits a hierarchical VAE of general-form autoregressive priors that permits parallel coding. Using our framework, we outperform all other VAE-based compression approaches with fewer latent variables and a comparable number of neural network evaluations. We further illustrate the effectiveness of our architecture by training a small model which is able to outperform similar VAE approach Bit-Swap \cite{kingma2019bit} -- but with 100x fewer model parameters. 

\section{Related Work}\label{sec:related}
\textbf{Compression with VAEs} can be separated into those \textit{with} \cite{kingma2019bit, townsend2019hilloc} and those \textit{without} \cite{mentzer2019practical} stochastic posterior sampling (the latter uses a discrete distribution of one symbol, assumed to have a probability of one). 
Whilst obtaining theoretically superior compression ratios, approaches adopting stochastic posteriors, such as HiLLoC \cite{townsend2019hilloc} and Bit-Swap \cite{kingma2019bit}, must entropy code using derivatives of the bits-back argument \cite{hinton93keeping}. These approaches, under the umbrella of bits-back ANS (bb-ANS), require access to an \textit{initial bitstream}. Although it is possible to amortize the cost of the initial bitstream across a large dataset, single and parallel data compression has presently proven challenging (see Section \ref{sec:bbans}). In HiLLoC \cite{townsend2019hilloc} the authors propose use of a conventional codec to compress and send parts of a dataset, which is then used as the basis for the initial bitstream. Whilst this partially solves some of the coding challenges, it requires the implementation of sub-optimal traditional techniques and still does not permit practical coding of a single image.  In contrast, our approach avoids all of these challenges, requiring fewer latent variables, for a negligible additional bit cost. Similarly, approaches such as L3C \cite{mentzer2019practical} that leverage deterministic posteriors can also avoid the challenges associated with bb-ANS with use of arithmetic (or adaptive arithmetic) coding (AC) \cite{10.1145/214762.214771}. Closely related approach RC \cite{mentzer2020learning} -- which can be loosely thought of as a VAE -- uses a lossy compressed image as a de facto latent variable to condition the data distribution. However these techniques pay for their comparative practicality with a penalty in compression ratio as they must explicitly code the joint distribution of data and latent variables (see Section \ref{sec:determ}). 
Since this work focuses on VAEs based codecs, we refer readers to \cite{hoogeboom2019integer,berg2020idf++, zhang2021ivpf, zhang2021iflow} and references therein for compression with alternative deep generative models.

\begin{figure*}[ht]
\centering
\includegraphics[width=.90\textwidth]{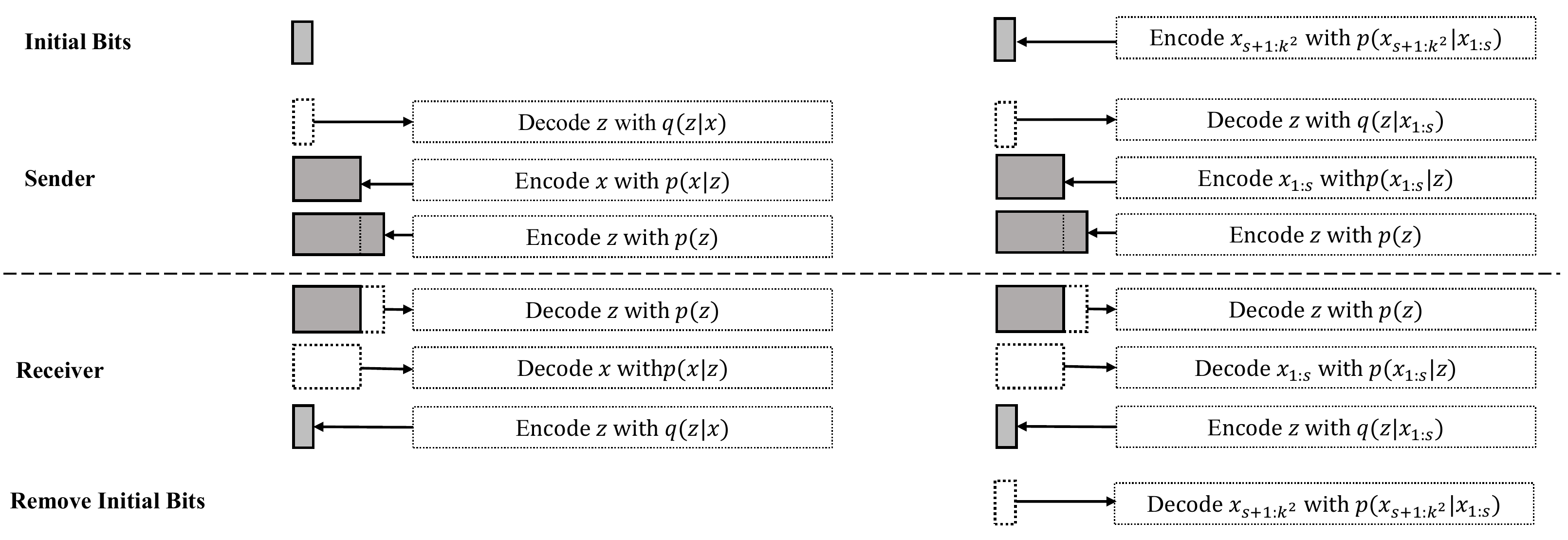}
\caption{Left: The bb-ANS coding framework as discussed in Section \ref{sec:bbans}. Right: The coding framework of ArIB as proposed in Section \ref{sec:arib}. Here we nest a bb-ANS coder inside of a block-based autoregressive structure that removes the need for initial bits.}
\label{fig:bbans}
\end{figure*}

\textbf{Autoregressive Models} are a popular means to extend the independence assumptions of fully factorised models to high-dimensional multivariate densities. They are popular as both stand-alone models \cite{van2016pixel, salimans2017pixelcnn++} or in application with VAEs \cite{NEURIPS2018_53edebc5, pmlr-v70-reed17a, gulrajani2016pixelvae}. In their most computationally expensive forms, such as PixelCNN++ (and variants) \cite{salimans2017pixelcnn++}, they posit a pixel-by-pixel autoregression, which then codes in raster-scan order. Whilst of broad academic interest, the $\mathcal{O} (n^2)$ time complexity makes them prohibitive for application in data compression. One proposed solution to this problem has been to parametrise the priors of VAEs with autoregressive densities. Here probability estimation proceeds by combining hierarchical latent representations with decoded autoregressive context. Supplementing autoregressive components with auxiliary latent features permits causality restrictions that reduce time complexity, without greatly diminishing performance. These restrictions include channel-wise autoregressions \cite{DBLP:conf/icip/MinnenS20}; independent, block-based models \cite{pmlr-v70-reed17a}; ``checkerboard" context \cite{Zhang_2020_ACCV}; and small neural networks \cite{gulrajani2016pixelvae}, amongst others. In fact, these restrictions are likely of dual purpose: combining powerful autoregressive models with VAEs will likely expedite posterior collapse (see Section \ref{sec:res3}) \cite{bowman2016generating, razavi2018preventing, gulrajani2016pixelvae, lucas2019don}. Like these techniques, our approach combines a VAE with a restricted autoregressive model. Our method can be thought as most similar to \cite{DBLP:conf/icip/MinnenS20} and \cite{Zhang_2020_ACCV}. 
Like the former, we perform a channel-wise autoregression, but do so after our autoregressive operator downsamples the data tensor. As such, our causality more closely resembles that of \cite{Zhang_2020_ACCV}. However, in contrast to the authors of \cite{Zhang_2020_ACCV}, who enforce their causality with a binary mask, we do so using our sub-pixel convolution. This precipitates a greater degree of parallelism and presents the flexibility to efficiently recover a number of causal dependency schemes, such as PixelCNN++.

\section{Background}\label{sec:background}
Suppose access to a dataset of size $n$, $\{x_1, x_2, \dots, x_n\}$, drawn from some intractable $p_{\text{data}}({x})$ that we wish to compress. In order to achieve this, we introduce a discrete probability distribution $p({x})$ that, in combination with \textit{entropy coding}, requires a \textit{codelength} of $\sum_{i=1}^n -\log_2 p(x_i)$ bits to represent.  Ideally, $p({x})$ should closely resemble $p_{\text{data}}({x})$. In such a case, the average codelength  in the limit of $n \longrightarrow \infty$ is given by $\mathbb{E}_{p_{\text{data}}}\left[ -\log_2 p(x)\right] \longrightarrow H({x})$, where  $H({x})$ is the entropy of the data. Here, the compression scheme is said to be \textit{optimal} under Shannon's source coding theorem \cite{10.1145/584091.584093}.

\subsection{Bits-back ANS for VAEs} \label{sec:bbans}
As discussed in Section \ref{sec:intro}, variational autoencoders (VAEs) are one popular approach to estimating $p_{\text{data}}({x})$, which defines a latent variable model such that
\begin{align}
    p({x}) = \int p({x}, {z}) d{z} = \int p({x}|{z})p({z})d{z},
\end{align}
where $p({z})$ is the \textit{prior} distribution over latent variable ${z}$. 
As $p({x})$ is normally intractable, VAEs introduce an approximate \textit{posterior} $q(z|x)$, which is optimised to maximise a lower bound on the marginal evidence, the Evidence Lower Bound (ELBO)

\begin{align}
\log p(x) \geq & \mathcal{L} = \mathbb{E}_{q} \bigg[ -\log q(z|x)+\log p(x|z)p(z)\bigg], \label{eq:ELBO}
\end{align}
where low-variance estimates of the expectation in \eqref{eq:ELBO} are via Monte-Carlo integration and the \textit{reparametrization trick} \cite{kingma2013auto}. 

Entropy coding requires explicit probabilities of data symbols.
However, in VAEs, the model is factorised into a prior and a likelihood, and therefore does not allow direct coding of the data.
To remedy this, several authors have proposed coding variants of bb-ANS \cite{townsend2019practical, kingma2019bit, townsend2019hilloc}. 
This process is outlined as follows, which, without loss of generality, we describe for a model with a single latent variable.
During compression, one decodes $z$  from some \textit{auxiliary} initial bits with $q(z|x)$; encodes $x$ with $p(x|z)$; and encodes $z$ with $p(z)$ to obtain the complete bitstream.
In the decompression stage, one decodes $z$ from the bitstream with $p(z)$; decodes $x$ with $p(x|z)$; encodes $z$ with $q(z|x)$ and thus returns the initial bits (hence \textit{bits-back} coding).
This technique is visualised in Fg. \ref{fig:bbans} Left.

The first decoding step common in existing bb-ANS codecs requires access to an \textit{initial bitstream}.
This requirement leads to several disadvantages when compared to AC.
Firstly, while the initial bits are returned \textit{after} decompression, the same bits are occupied and not readily readable \textit{beforehand}.
Secondly, although we can amortize the cost of the initial bits across a large dataset by chaining the compressed data, access to any given data point requires decompression of all data points posterior to the one of our target in the original data sequence. 
As such, compression of a \textit{single} image carries substantial overhead -- and, by extension, so too do parallel coding implementations.

\subsection{Deterministic vs Stochastic Posterior Sampling} \label{sec:determ}
Within VAEs based lossless compression, the impracticalities associated with bits-back coding are not the only choice. Indeed, approaches that use ``deterministic" posterior sampling \cite{mentzer2019practical} may eschew bb-ANS for AC. This approach is almost ubiquitous in the lossless codecs of lossy approaches \cite{ball2018variational, NEURIPS2018_53edebc5, cheng2020image, DBLP:conf/icip/MinnenS20}, where
eminently parallelizable, low-latency codecs
are especially preferable (e.g. streaming media).
We note that when using deterministic posterior sampling, the likelihoods associated with the prior in Eq. \eqref{eq:ELBO} trivially cancel to zero, such that the objective resolves to maximum likelihood estimation of the joint distribution (see e.g. \cite{ball2018variational}).

Whilst gaining notable practical coding advantages, in sacrificing a stochastic posterior one also sacrifices the ability to minimize the cost of sending latent variables. 
To offset this limitation, models will repeatedly downsample (RDS) the number of symbols available to latent representations between each layer. 
Whilst this limits posterior expressiveness, there is little experimental evidence to support how much this matters in practice. 
In addition, models with stochastic posteriors require large $L$ to excel (where $L$ is the number of latent variables), which hinders run-time.

\begin{figure}[!tbp]
  \centering
    \includegraphics[width=0.49\textwidth]{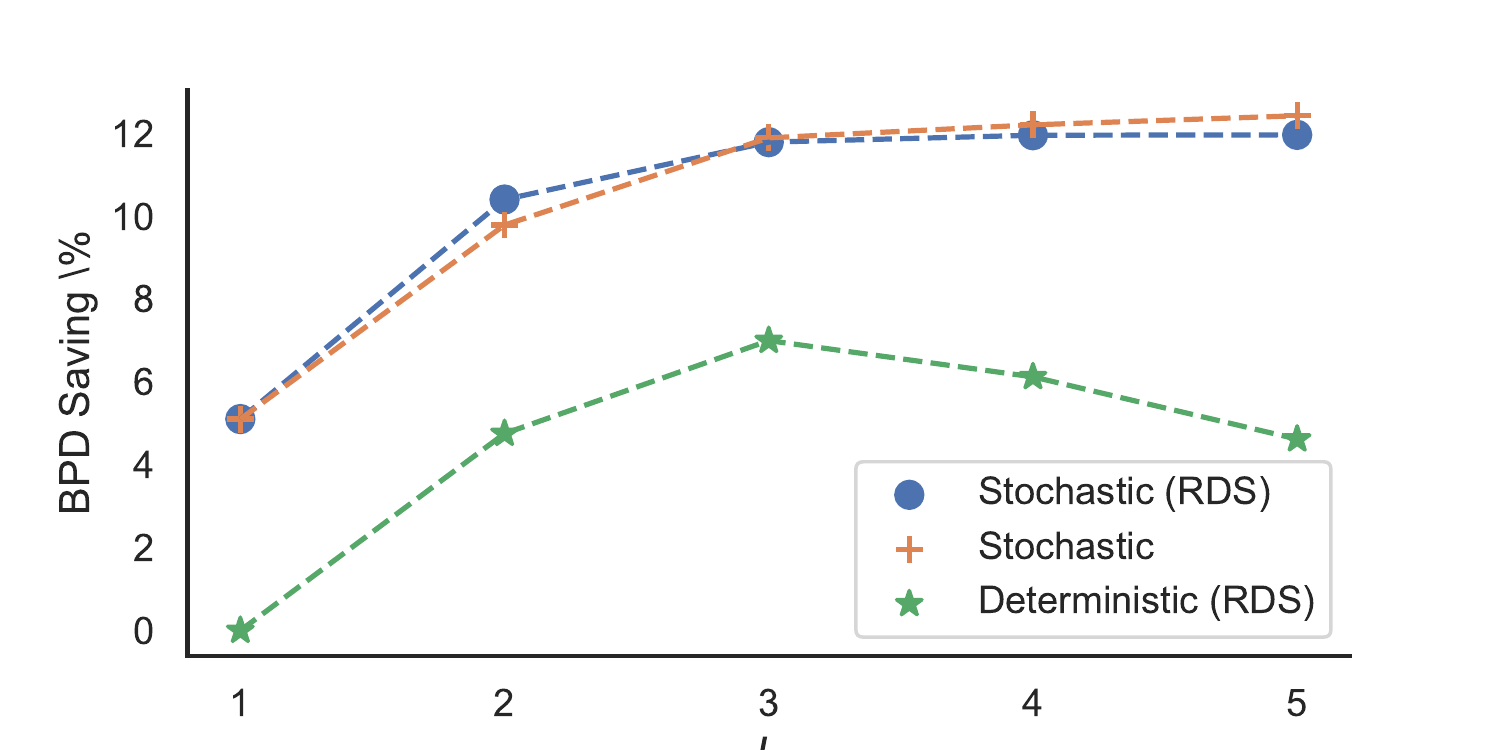}
    \caption{Average BPD savings against the number of latent variables ($L=1,...,5$) used in three VAE lossless compression models. The baseline is the deterministic posterior with $L=1$ where the BPD saving is zero. Results are presented as an average for the three approaches trained across CIFAR10, ImageNet32 and ImageNet64.}
    \label{fig:exp_1}
\end{figure}

To that end, we display the results of a simple experiment in Fg. \ref{fig:exp_1} designed to investigate this difference further. 
Here we train three VAEs across CIFAR10, ImageNet32 and ImageNet64: two with stochastic posteriors (one with and one without RDS) and a deterministic posterior (with RDS). The architectures in each model are identical (with the exception of downsampling operations) and we quantify compression performance in \textit{bits per dimension} (BPD). Further experimental details can be found in the Appendix. Here we observe that, even with $L\leq 3$, both stochastic posteriors outperform by $\sim 5 \%$. This difference grows as $L$ increases -- but it does not matter if RDS is used (at least, for $L\leq 5$). This result is of important consequence: the best current approach to avoid the impracticalities of bb-ANS (ie. using a deterministic posterior) carries a 5\% BPD penalty. For single-image compression with stochastic posteriors, the initial bits required would typically be much larger than this. Likewise, unless extending to a deep hierarchy of latent variables, RDS seems like a compute-efficient choice that does not limit performance.

\section{Method}

Our method posits a hierarchical VAE where we parameterize the priors using an autoregressive factorisation. We begin by defining a lossless downsampling convolution operator, before describing its application to density estimation using both \textit{weak} and \textit{strong} autoregressive models. We then describe how this autoregressive structure can be leveraged to avoid many of the challenges associated with bb-ANS \textit{without} sacrificing the performance of stochastic posteriors. Finally, we describe how these contributions can be combined to form our SHVC codec.

\begin{figure*}[!tbp]
\centering
\includegraphics[width=.750\textwidth]{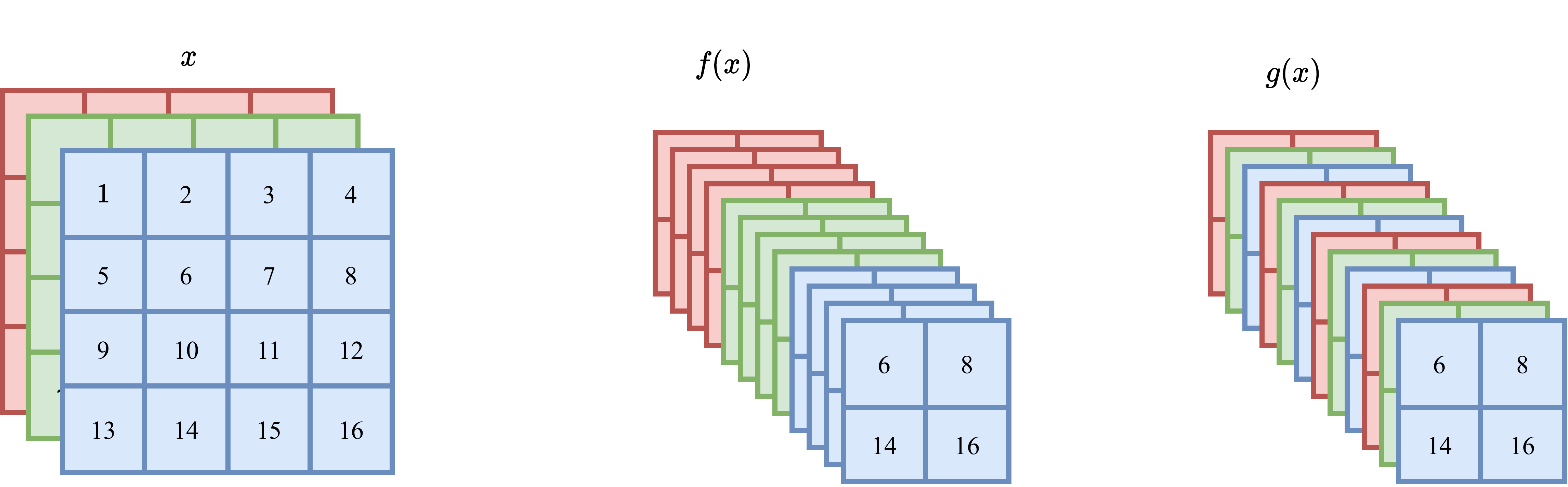}
\caption{Left: a $3 \times 4\times 4$ input RGB image. Centre: the image $x$ downsampled using the convolution operator of \eqref{eq:conv_op} with $k=2$. Right: the image $x$ downsampled using the convolution operator of \eqref{eq:conv_op2} with $k=2$.}
\label{fig:ps}
\end{figure*}

\subsection{Autoregressive Sub-Pixel Convolutions}\label{sec:aspc}

The space-to-depth and depth-to-space transformations are popular operations across image analysis, from generative modelling \cite{hoogeboom2019integer, berg2020idf++} to super-resolution \cite{shi2016real}. They define adjacent operations for efficient up and downsampling transformations by folding spatial dimensions into channel dimensions -- and vice versa. Unlike learned operations, they greatly reduce computational complexity, allowing for greater parallelism by losslessly moving computation (and data) into the channels. Indeed, these operations have become an essential component in papers seeking real-time execution (e.g. \cite{shi2016real, waveone2021elf, cortinhal2020salsanext, liu2018deep}). Specifically, given a tensor of $C$ channels, $H$ height and $W$ width, we define the space-to-depth and depth-to-space transformations, $f$ and $f^{-1}$, such that  
\begin{align}
    f&: \mathbb{R}^{C \times H \times W} \longrightarrow \mathbb{R}^{Ck^{2} \times \frac{H}{k} \times\frac{W}{k}}, \\
    f^{-1}&: \mathbb{R}^{C \times H \times W} \longrightarrow \mathbb{R}^{\frac{C}{k^{2}} \times Hk \times Wk}, 
\end{align}
where $k$ is the scale factor.

As described in \cite{shi2016real}, these operations can be efficiently performed using \textit{sub-pixel convolutions}, which are referred to as \textit{pixel unshuffle} and \textit{pixel shuffle}. In particular, their space-to-depth transformation, pixel unshuffle, is performed using a $k$-stride depthwise convolution where the $n^{th}$ element of $Ck^2$ $k \times k$ filters has one non-zero element such that

\begin{equation}
   K_{h, w}^{(n)}  =  \begin{cases} 
      1 & \text{if }h = \floor{n\mathbin{/}k}\, \text{mod} \, k, \, w=n \, \text{mod} \, k \\
      0 & \text{else}
   \end{cases}, \label{eq:conv_op}
\end{equation}
where $h, w$ are the indices over spatial dimensions. The result of this operation is visualised in Fg. \ref{fig:ps} Centre. 

Defining a channel-wise autoregression over the resulting tensor would posit a checkerboard autoregressive structure over each of the channels in the original tensor, sequentially. However, as identified in PixelCNN++ \cite{salimans2017pixelcnn++}, sub-pixels in adjacent channels, sharing the same \textit{spatial} location in the original tensor, have high correlation and therefore do not require complex models to describe the dependency structure. As such, the authors of PixelCNN++ use a linear model predicted by a single network evaluation, conditioned on decoded \textit{context}, to define the joint distribution across channels. In this way, they obfuscate the need for separate RGB network evaluations. (We note that in our setting, context refers to previously decoded pixels in either the current or previous hierarchical latent variable.) From henceforth what we refer to as a \textit{weak autoregression}, is then defined similarly to \cite{salimans2017pixelcnn++} according to

\begin{align}
p\left(x_{0:C, h, w}| D\right) = p\left(x_{0, h, w} | D\right)\prod_{c=1}^{C}p\left(x_{c, h, w}| x_{< c, h, w}, D\right) \label{eq:weak_ar_1}
\end{align}
where $D$ is the decoded context and $p$ is some parametric probability mass function (pmf), obtained via integrating a probability density function (pdf) over discretization bins,  with mean at channel $c$ location $h, w$ given by
\begin{equation}
    \mu_{c, h, w} = \alpha_{c, h, w} + \sum_{i=0}^{c-1} \beta_{c, h, w}^{(i)} x_{i, h, w}. \label{eq:weak_ar_2}
\end{equation}

Here $\alpha$ and $\beta$ are scalars predicted for all channels and spatial locations by a single network evaluated on decoded context, and $i$ is the index over channels in decoded context such that $\beta_{c, h, w}^{(i)}$ is the scalar for prediction of the mean associated with pixel at channel $i$, spatial location $h, w$.

Inspired by this, we introduce a new space-to-depth convolution such that the resulting autoregression is alternatively re-ordered into $k^2$ \textit{sub-blocks} of $C$ channels each. Crucially, the resulting channels in each sub-block share the same spatial index allowing application of the autoregression detailed in \eqref{eq:weak_ar_1} and \eqref{eq:weak_ar_2}. 
We note that should $k=H=W$ we return an equivalence to the per-pixel autoregression of PixelCNN++ but perform an autoregression exclusively in the channel dimension. Likewise should $k<H$, we define a block-based context model in raster scan order where, unlike MS-PixelCNN, adjacent blocks are dependent.  

To achieve our desired downsampling operation, which we denote by $g(\cdot)$, we expand the depthwise convolutions of \eqref{eq:conv_op} into regular three-dimensional kernels where the $n^{th}$ of $Ck^2$ $C \times k \times k$ filters has one non-zero element such that
\begin{equation}
   K_{c, h, w}^{(n)}  = 
   \begin{cases} 
      1 & \text{if }c = n \,\, \text{mod} \,\, C, \, h = \floor{n\mathbin{/}Ck}\, \text{mod} \, k, \, \\
      & \, \, \, \, w=\floor{n\mathbin{/}C} \, \text{mod} \, k \\
      0 & \text{else}
   \end{cases}. \label{eq:conv_op2}
\end{equation}
We further visualise this operation in Fg. \ref{fig:ps} Right. The resulting density of the downsampled tensor for spatial location $h, w$ is then given by 
\begin{align}
&p\left(g\left(x; k\right)_{0:Ck^2, h, w}| D\right) = \prod_{i=0}^{k^2} \bigg[p\left(g\left(x; k\right)_{iC, h, w} | D\right) \times \nonumber\\
& \; \; \; \; \; \; \; \; \; \; \; \; \; \; \; \times \prod_{j=iC+1}^{(i+1)C -1}p\left(g\left(x; k\right)_{j, h, w}| g\left(x; k\right)_{<j, h, w}, D\right) \bigg]\label{eq:weak_ar_3},
\end{align}
where $i$ is the index over sub-blocks (i.e. a strong autoregression evaluated using neural networks).


\paragraph{Masked 3D Convolutions}
Whilst we are restricted to $k^2$ evaluations per latent variable at inference time, the same does not have to be true during training. One efficient parallel training scheme is to use 3-dimensional convolutions applied to the downsampled tensors by expanding them into a $d \times Ck^2 \times H \times W$ volume \cite{pmlr-v48-oord16, mentzer2018conditional1}, where $d$ is some auxiliary dimension. Here we can apply zero-masking along the channel dimension of the kernels to enforce the causality condition, along with k-stride channel convolutions on the input. Full details are available in the Appendix.

\paragraph{Choice of Distribution}
For our choices of $p$ and $q$, we use a discretized mixture of logistic distributions for $x$ and a discretized univariate logistic distribution for all $z^{(l)}$ \cite{salimans2017pixelcnn++}. 
That is, given some mean $\mu$, scale $s$ and uniform discretization bin-width $b$, one can obtain the univariate pmf by integrating the logistic pdf over the discretiztion bin. 
For $x$, we typically use a mixture of 5 discrete logistic distributions as defined above.

\subsection{Autoregressive Initial Bits}\label{sec:arib}
As discussed in Section \ref{sec:bbans}, bb-ANS is able to achieve efficient codelengths, but can lead to several shortcomings.
Fortunately, our proposed autoregressive model naturally accommodates the possibility to bypass the auxiliary bits needed in other bb-ANS methods. We achieve this by exploiting the block-based autoregressive structure on the data variable. We outline this process below, which we refer to as \textit{autoregressive initial bits} (ArIB). 
Different from the models considered in existing VAE-based codecs, we remove the direct causality between the latent variable $z$ and some partition of the data variable $x$. In practice, we simply remove $z$ from $D$ in Eq. \eqref{eq:weak_ar_3} for the final $n$ sub-blocks in $p(x|z)$, along with the partition from $x$ in $q\left(z^{(1)}|x\right)$.
As a result, we factorise the likelihood as $p(x|z)=p(x_{s+1:k^2}|x_{1:s})p(x_{1:s}|z)$ with the approximate posterior as $q(z|x)=q(z|x_{1:s})$, where $s$ is our `split' index.
Instead of conducting the first step by decoding $z$ from $q(z|x)$, one can encode $x_{s+1:k^2}$ with $p(x_{s+1:k^2}|x_{1:s})$ and thus obtain the bitstream from which to decode $z$.
Then one decodes $z$ with $q(z|x_{1:s})$, encodes $x_{1:s}$ with $p(x_{1:s}|z)$ and encodes $z$ with $p(z)$.
At the decompression stage, one decodes $z$ with $p(z)$, decodes $x_{1:s}$ with $p(x_{1:s}|z)$, encodes $z$ with $q(z|x_{1:s})$ and decodes $x_{s+1:k^2}$ with $p(x_{s+1:k^2}|x_{1:s})$. We illustrate this technique in Fg. \ref{fig:bbans} Right.

For this approach to be valid, we require the satisfaction of two criteria:
\begin{enumerate}
    \item There exists some $s$, $k$ and $z$ such that imposing $\left(x_{s+1:k^2} \perp z | x_{1:s}\right)$ does not greatly hinder performance. 
    \item The \textit{entropy} of $p\left(x_{s+1:k^2}|x_{1:s}\right)$ and $q\left(\hat{z}|x_{1:s}\right)$, where $\hat{z}$ is the discretized analogue of $z$, should be such that $\mathcal{H}_{p(x_{s+1:k^2}|x_{1:s})} \geq \mathcal{H}_{q(\hat{z}|x_{1:s})}$. 
\end{enumerate}
In our experiments, we demonstrate that the performance costs associated with criteria one are negligible. Crucially, we demonstrate that it is both orders of magnitude less that initial bits required of vanilla bb-ANS and a parameterisation of our approach using deterministic posteriors. 

For criteria two, we formulate the optimization of \eqref{eq:ELBO} as a constrained problem subject to $\mathcal{H}_{p(x_{s+1:k^2}|x_{1:s})} \geq \mathcal{H}_{q(\hat{z}|x_{1:s})}$, where we estimate the respective expectations during training using Monte-Carlo integration. Whilst a variety of techniques from optimization theory may be applied, we found it sufficient to simply penalise \eqref{eq:ELBO} according to 
\begin{equation}
    \mathcal{L}_{pen} = \mathcal{L} + \lambda \max \left(0,  \mathcal{H}_{q(\hat{z}|x_{1:s})} - \mathcal{H}_{p(x_{s+1:k^2}|x_{1:s})} \right), \label{eq:constrained}
\end{equation}
where $\lambda$ is some Lagrange multiplier. We find that this further presents flexibility when choosing $s$, with a variety of choices yielding the same result. 

\begin{table*}[h]
  \label{table:results}
  \centering
\begin{tabular}{lcccc|ccc}\toprule
         &Compression Model& CIFAR10 & ImageNet32 & ImageNet64 & CLIC.mobile& CLIC.pro&DIV2K \\
\midrule
\textit{Generic}&PNG \cite{boutell1997png}     &5.71        & 5.87& 6.39 &3.90 & 4.00 & 3.09 \\
&FLIF \cite{sneyers2016flif}            & 4.19        & 4.19& 4.52&2.49 & 2.78 & 2.91 \\
&JPEG-XL \cite{alakuijala2019jpeg}  & 5.74 & 5.89& 6.39 &2.36 & 2.63 & 2.79 \\
\midrule
\textit{VAE-Based} 
&L3C \cite{mentzer2019practical}      & - & 4.76        & 4.42           &2.64 & 2.94 & 3.09 \\
&Bit-Swap \cite{kingma2019bit} & 3.82& 4.50 & - &  - & - & -\\
&HiLLoC \cite{townsend2019hilloc} &  3.56$^{\ddag}$ & 4.20$^{\ddag}$ & 3.90$^{\ddag}$ &  - & - & - \\
&\textbf{SHVC} & \textbf{3.16/3.41$^{\ddag}$} & \textbf{3.98} & \textbf{3.68/3.71$^{\ddag}$} & \textbf{1.96}$^{*}$ & \textbf{2.02}$^{*}$ & \textbf{2.57}$^{*}$\\
& \textbf{SHVC Lite}& \textbf{3.76} & \textbf{4.49 }& \textbf{4.16} &  - & - & - \\
\midrule
\textit{Flow-Based} &IDF \cite{hoogeboom2019integer} & 3.34/3.60$^{\ddag}$ & 4.18 & 3.90/3.94$^{\ddag}$ &  - & - & -\\
&IDF++ \cite{berg2020idf++} & 3.26 &  4.12 & 3.81 &  - & - & -\\
&LBB  \cite{ho2019compression}  & 3.12 & 3.88 & 3.70 &  - & - & -\\
&iVPF  \cite{zhang2021ivpf}& 3.20/ 3.49$^{\ddag}$&  4.03  & 3.75/3.79$^{\ddag}$ &2.39$^*$ & 2.54$^*$ & 2.68$^*$\\
& iFlow \cite{zhang2021iflow} & {3.12}/3.36$^{\ddag}$  & {3.88} & {3.70}/{3.65}$^{\ddag}$ &{2.26}$^*$ & {2.44}$^*$ & {2.57}$^*$ \\
\bottomrule
\end{tabular}
  \caption{Compression results in BPD for SHVC and other popular codecs across three low-resolution datasets and three full-resolution datasets. Lower is better. Here $^{\ddag}$ and $^*$ denote models trained on ImageNet32 and ImageNet64, respectivley.}
\label{table:results}
\end{table*}

\subsection{Split Hierarchical Variational Compression}

SHVC formulates a hierarchical VAE built from the components described above. Here we partition the latent variable into a simple disjoint hierarchy of $L$ layers, such that $z =\{z^{(1)}, . . ., z^{(L)}\}$. We define the prior and posterior according to
\begin{align}
    p\left(x, z^{(1:L)}\right) &=  p\left(x|z^{(1)}\right) p\left(z^{(L)}\right) \prod_{i=1}^{L-1} p\left(z^{(i)}| z^{(i+1)}\right), \label{eq:prior} \\
    q\left(z^{(1:L)}|x\right) &= q\left(z^{(1)} | x\right) \prod_{i=1}^{L-1} q\left(z^{(i+1)}| z^{(i)}\right),
\end{align}
where we parameterise every conditional density in Eq. \eqref{eq:prior} as per Eq. \eqref{eq:weak_ar_3}.
While this factorisation naturally fits the coding scheme proposed in Bit-Swap \cite{kingma2019bit}, we additionally introduce a local \textit{reverse} encoding to accommodate the autoregressive structure for factors in Eq. \eqref{eq:prior}.
In more detail, for the encoding of $z^{(i)}=[z^{(i)}_1, ..., z^{(i)}_{k^2}]$ with $p(z^{(i)}| z^{(i+1)})$, one needs to encode in the reserved order of $z^{(i)}_{k^2}, ..., z^{(i)}_1$, to accommodate the first-in-last-out nature of ANS based codecs.

For purposes of experimentation, we define two versions of our model: one with and one without the dependency structure permitting ArIB. From henceforth we shall refer to these models as SHVC and SHVC-ArIB, respectivley. 
For SHVC, one can encode $x$ with $p(x|z^{(1)})$, along with other variables in Eq. \eqref{eq:prior} as discussed above.
For SHVC-ArIB, one performs encoding and decoding for $x$ as discussed in Section \ref{sec:arib}; and applies local reverse encoding for slices in $x_{s+1:k^2}$ and $x_{1:s}$, respectively. We note that the only difference in SHVC-ArIB is that, whilst $p(x_{s+1:k^2} | x_{1:s})$ and $p(x_{1:s}|z^{(1)})$ are both modelled using \eqref{eq:weak_ar_3}, the former evidently omits $z^{(1)}$ from $D$. In addition, we restrict the posterior such that $q(z^{(1)}|x_{1:s})$. We visualise the overall architecture along with the coding scheme for SHVC-ArIB in the Appendix. 
\section{Experiments}
In this Section, we perform a series of experiments to evaluate the effectiveness of compression with SHVC. We begin by discussing the architecture and training details in Section \ref{sec:training}. 
In Section \ref{sec:res1}, we evaluate compression performance in terms on both low and full-resolution images. Here we additionally evaluate the effect of the ArIB constraints on compression performance and assess inference speed. Finally, in \ref{sec:res3} we perform a series of ablation studies.

\subsection{Architecture and Training Details} \label{sec:training}

As discussed in Section \ref{sec:background}, we observe an efficient trade-off between inference speed and performance where we repeatedly downsample and limit the number of latent variable layers. Inspired by this, we apply repeated downsampling, using three and four latent variables when training $32 \times 32$ and $64 \times 64$ images, respectively. With these settings, we require 3 and 16 network evaluations for posterior and prior inference on CIFAR10. In contrast, leading VAE approach HiLLoC \cite{townsend2019hilloc} requires 24 prior \textit{and} posterior evaluations. 

For all of our experiments, we set $k=2$ in our autoregressive model. Whilst this choice is discussed more extensively in Section \ref{sec:res3}, we found the $k=2$ presented an effective compromise between inference speed and representational power. Indeed, increasing $k$ can  often result in \textit{worse} performance. We observe that increasing $k$ increases the possibility of posterior collapse \cite{bowman2016generating, razavi2018preventing, gulrajani2016pixelvae, lucas2019don}, making training a hierarchy of powerful autoregressive models challenging. We further detail comprehensive architectures for all of our experiments in the Appendix.

\subsection{Compression Performance} \label{sec:res1}

\paragraph{Low-Resolution Images}
We begin by testing our method on three toy datasets: CIFAR10, Imagenet32 and Imagenet64. We compare our method against leading approaches from traditional codecs, normalizing flows and VAEs. (Given the time complexity associated with per-pixel autoregressive factorisations, we follow the broader compression community and eschew them from our comparisons). As an ad hoc test of generalizability, we additionally follow the authors of \cite{townsend2019hilloc, zhang2021ivpf, hoogeboom2019integer, zhang2021iflow} by training a model on Imagenet32 and testing it across all other datasets. We present our results in terms of bits per dimension (BPD), which we display in Table \ref{table:results}.

\paragraph{Full-Resolution Images}
We further compare our approach against full-resolution algorithms, i.e., L3C \cite{mentzer2019practical}. Here we follow \cite{zhang2021ivpf} and adopt our Imagenet64 model using a patch-based evaluation protocol in which images are cropped to $64\times 64$. From Table \ref{table:results}, we demonstrate  reliable out-performance of every other considered codec. Further, when comparing the performance of SHVC to e.g. iFlow across small and full-resolution images we interestingly note that difference in out-performance becomes greater. Indeed, we hypothesise that autoregressive "context" becomes an increasingly important inductive bias as the resolution of the image increases -- and likewise the extent of the spatial redundancy.

\begin{figure*}[h]
\centering
\includegraphics[width=.80\textwidth]{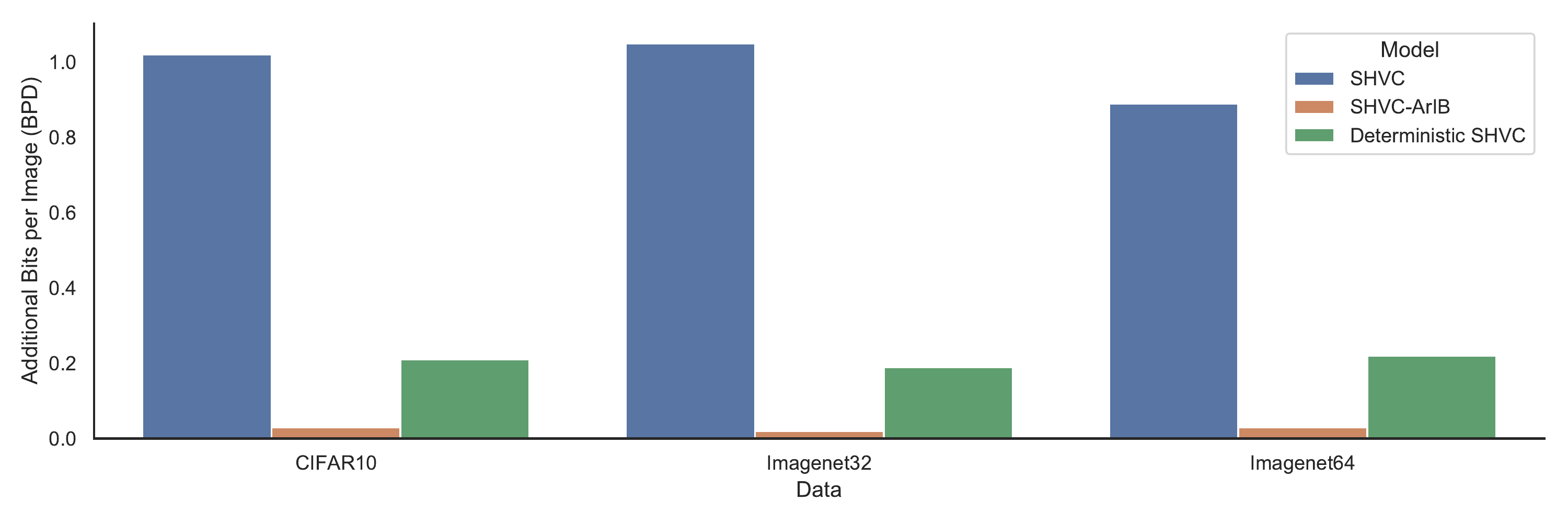}
\caption{Additional bits per single image compressed across SHVC, SHVC-ArIB and Deterministic SHVC. The overhead in SHVC comes from initial bits, whilst other models incur a performance cost.}
\label{fig:p_tests}
\end{figure*}

\paragraph{ArIB}
ArIB poses non-trivial constraints on the data variable, such that a sufficiently large partition should be: 1. conditionally independent of latent variables; 2. encoded with an entropy larger than that with which the first latent variable is decoded with. Using the objective of Eq. \eqref{eq:constrained}, we train models capable of single-image compression: SHVC, SHVC-ArIB and a parameterisation of SHVC with deterministic posteriors using AC (henceforth known as \textit{Deterministic} SHVC). We display results across CIFAR10, Imagenet32 and Imagenet64 in Fg. \ref{fig:p_tests}, and quantity our results as additional bits on a per-image basis against SHVC. In more detail, the overhead in SHVC comes from initial bits, whilst other models incur a performance cost. Here we see that our ArIB adds minimal additional bits -- less than the number in SHVC by a factor of $\sim$20. Crucially, our approach also outperforms Deterministic SHVC, which we hope may serve as motivation to adapt the codec to lossy compressors.

\begin{table}[h]
  \centering
\begin{tabular}{lllll}
\toprule
                    & \textbf{SHVC} & Bit-Swap & HiLLoC & IDF \\
                    \midrule
BPD                 &   \textbf{3.18}   &     3.82     &     3.32  & 3.34 \\
Time (s) &   \textbf{4.63}  &    5.86  &   10.20   &  20.58\\
\bottomrule 
\end{tabular}
\caption{Inference time and BPD across CIFAR10. Inference time measures evaluation of 10,000 test images with a batch size of 100. }
\label{table:runtime}
\end{table}
\vspace{-4.00mm}

\paragraph{Inference Speed}
To better evaluate our approach in the context of VAEs, we compare SHVC against popular publicly available compression models, HiLLoC \footnote{In contrast to SHVC and Bit-Swap, HiLLoC is implemented in TensorFlow. As such, HiLLoC is likely advantaged in terms of run-time environment.} \cite{townsend2019hilloc}, Bit-Swap \cite{kingma2019bit} and Integer Discrete Flow (IDF) \cite{hoogeboom2019integer} \footnote{We note follow-up paper IDF++ \cite{berg2020idf++} has no publicly available code but uses largely the same architecture as IDF. As such IDF++ inference times can be loosely inferred from IDF speeds.}.
We report achieved BPDs in Table \ref{table:runtime} and measure inference time in seconds (s) to evaluate the 10,000 CIFAR10 test images with a batch size of 100. Here we observe that our model is faster than HiLLoC for a lower BPD and achieves lower BPD for the same speed as Bit-Swap. We further note that, unlike both considered approaches, we are able to easily achieve parallel coding with minimal overhead.

\subsection{Ablation Studies} \label{sec:res3}
\begin{figure}[!tbp]    
    \centering
    \includegraphics[width=0.48\textwidth]{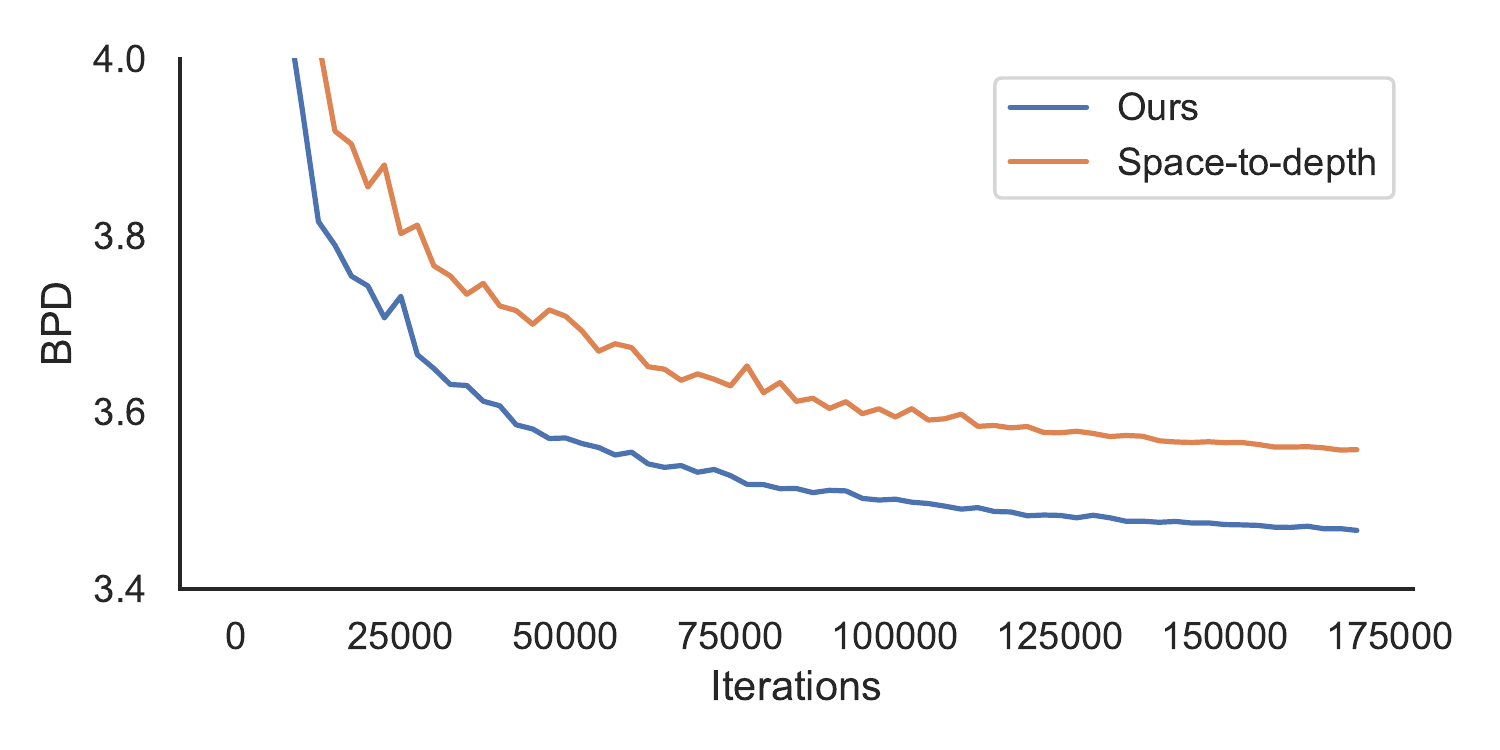}
    \caption{The theoretically minimal BPD against iteration, comparing the common space-to-depth operation vs ours.}
    \label{fig:ds_op}
\end{figure}
\paragraph{Space-to-depth operation}
One alternative in our approach is to replace our downsampling operator of \eqref{eq:conv_op2} with the usual space-to-depth transformation described in Eq. \eqref{eq:conv_op}. However, as discussed, this would instead define a weak autoregressive property over spatially adjacent pixels, channel-by-channel. In Fg. \ref{fig:ds_op} we demonstrate the differences resulting from the choice of spatial downsampling transformation by training two models on CIFAR10. Here we observe that our convolutional operator provides non-trivial benefits over the vanilla space-to-depth transformation. 

\begin{figure}[tbp]
    \centering
    \includegraphics[width=0.48\textwidth]{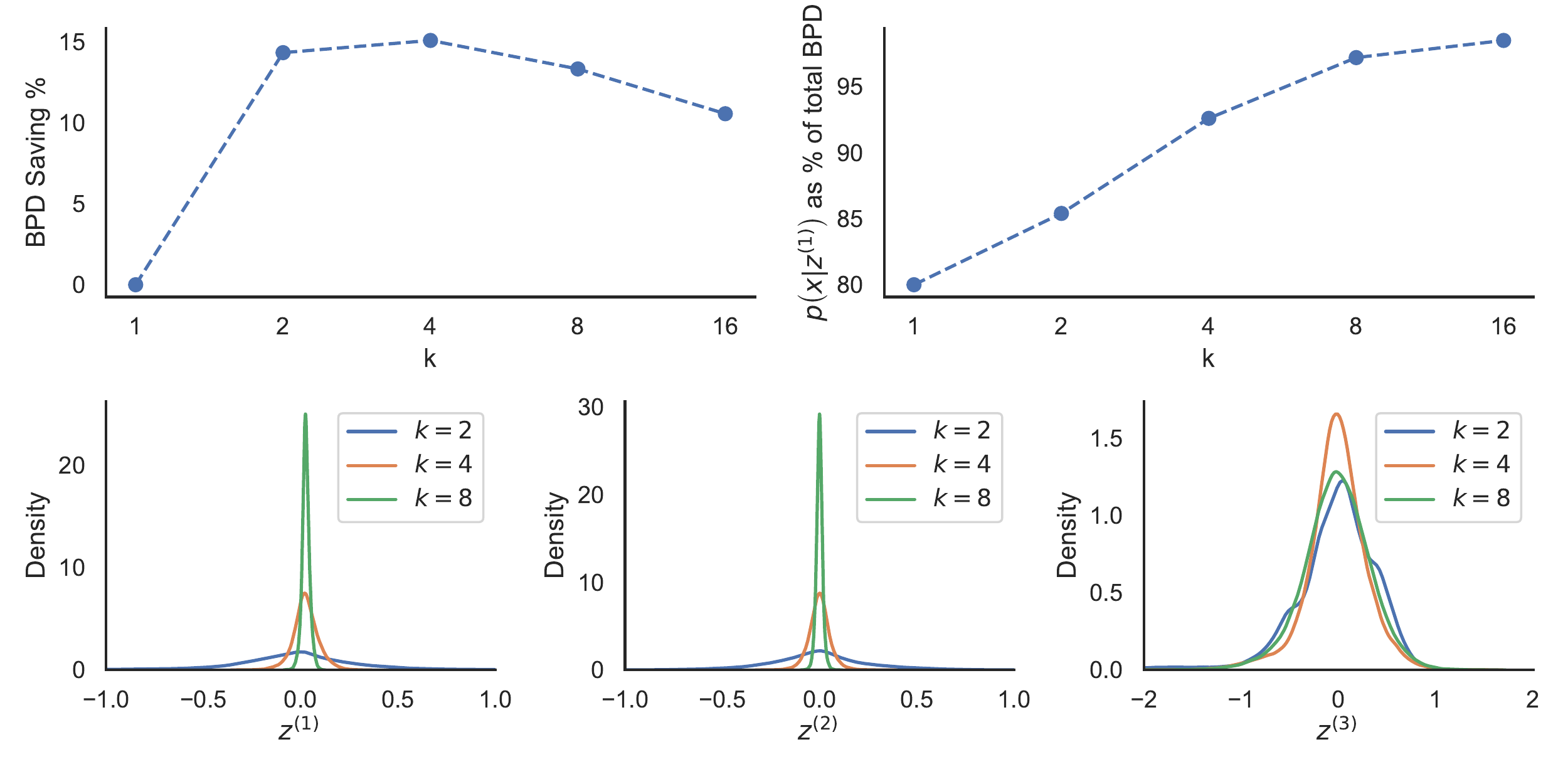}
    \caption{Top: (left) BPD savings against the choice of $k$ (the baseline is $k=1$), and (right) the reconstruction loss as a percent of the total BPD. Bottom: marginal densities of latent variables for different $k$.}
    \label{fig:pc_1}
\end{figure}

\paragraph{Choice of $k$}
As discussed, one important hyper-parameter choice is that of $k$. As $k$ grows, the prior becomes more powerful but the time complexity grows. In Fg. \ref{fig:pc_1} we visualise the effect of increasing $k$ on the compression ratio, which is displayed as an average across three models trained on CIFAR10, ImageNet32 and ImageNet64. As discussed, we note that performance of the model peaks at $k=4$, before becoming worse as $k$ increases. This non-intuitive behaviour can be better explained further in Fg. \ref{fig:pc_1}, where we see evidence of a posterior collapse common in hierarchical VAEs -- especially those with hierarchical autoregressive priors \cite{bowman2016generating, razavi2018preventing, gulrajani2016pixelvae, lucas2019don}.

\section{Conclusion}
We have proposed and evaluated a new VAE model for data compression. SHVC is able to outperform existing VAE approaches in terms of both speed and compression ratios. Additionally, unlike competing approaches using bb-ANS, SHVC-ArIB is able to support parallel coding with minimal overhead. As such, we believe ArIB represents the most promising means to convert a VAE into a lossless codec. Motivated by this, one promising area of future work could consider the application of SHVC to lossy compression, which has traditionally ignored bits-back schemes.

Whilst our approach does not introduce any new negative societal consequences, we remain acutely aware of the issues precipitated by this area of research. These concerns are two-fold. Firstly, improving the capacity to store data could greatly increase the amount of non-essential personal data held by third-parties. This may be viewed as a challenge on broader digital liberties. Secondly, any class of generative model trained on sensitive data will learn to closely approximate the distribution of that data. As such, the model itself might extend to malignant use-cases beyond its intended purpose., e.g. classifiers.
{\small
\bibliographystyle{ieee_fullname}
\bibliography{egbib}
}

\newpage
\makebox{}
\newpage
\appendix

\section*{A.1 Experimental Details for Section 3.2}

For both the encoders and decoders, we use four fully-connected two-dimensional convolutional layers with 128 channels and a 3x3 kernel. We additionally use weight-normalization at each layer and PReLU activation units. Where relevant, we downsample with 2-stride convolutions at the third convolution, and upsample using a transposed convolution at the third convolution. 

For training, we use the Adam optimizer with default learning settings and an initial learning rate of $5\times 10^{-4}$. We exponentially anneal this learning rate to $1 \times 10^{-5}$ during training.

Unlike stochastic posterior sampling, where we can train with continuous latent variables because discretization schemes cancel across distributions, deterministic posterior sampling requires discretization during training. Because the discrteization operation (i.e. rounding) is not differentiable, we adopt the popular technique of adding uniform noise during training, such that our discretized latent variable is defined by
\begin{equation}
    \hat{z}^{(l)} = z^{(l)} + \epsilon, \quad \epsilon \sim \mathcal{U}\left(-\frac{\delta}{2}, \frac{\delta}{2} \right),
\end{equation}
where $\delta$ is the uniform discretization bin and $\mathcal{U}$ represents a uniform distribution. In practice, we take $\delta=1$.

\section*{A.2 Masked 3D Convolutions}
For large $k$ it becomes impractical to train using two-dimensional convolutions. Doing so typically necessitates a serial scheme across data partitions at a given latent variable. One approach to train our models in parallel is to use masked three-dimensional convolutions. We achieve this by expanding our downsampled data tensors into a $d \times Ck^2 \times H \times W$ volume, where $d$ is some auxiliary dimension. 

\begin{figure}[h]
\centering
\includegraphics[width=.4\textwidth]{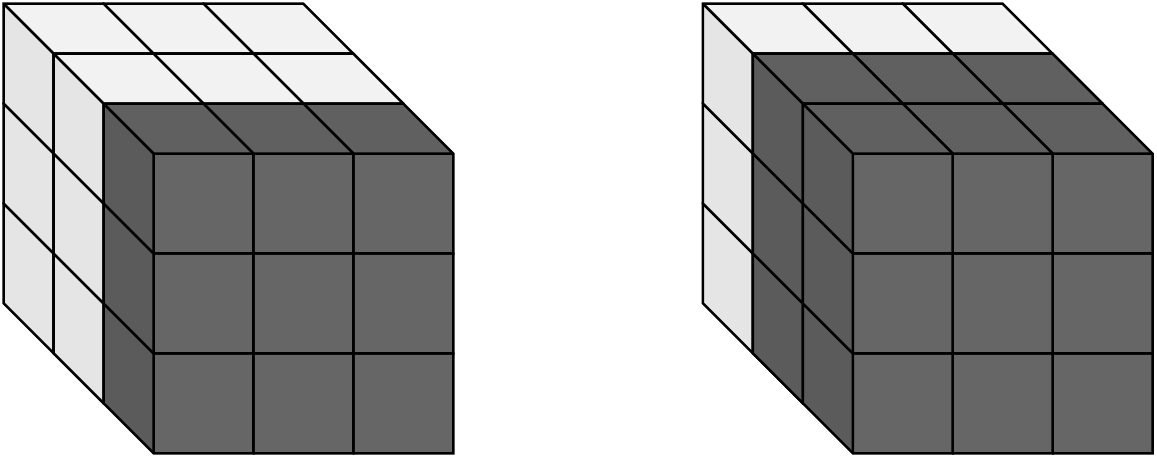}
\caption{A $3 \times 3 \times 3$ kernel masking strategies. Mask type `B' left; type `A' right. Dark grey elements indicates zeros; light elements indicate ones. Masks are applied as point-wise multiplications to the kernel.}
\label{fig:mask}
\end{figure}

In order to retain the causality constraints, we build our approach of two steps:
\begin{enumerate}
    \item We use an \textit{off-center convolution} of stride $C$ to enforce the autoregressive structure \textit{within} sub-blocks. We define this operation as one that concatenates a zero-tensor of dimension  $d \times C \times H \times W$ to the data variable along the auxiliary dimension and then applies the convolution as described. The result of applying this convolution is a $f \times k^{2} \times H \times W$ tensor, where $f$ are the output channels of the convolution.
    \item We then apply repeated \textit{masked} three-dimensional convolutions to the output of the the off-center convolution. To enforce the causality constraint between sub-blocks we apply a point-wise mask to the kernels prior to convolution. We define two types of masks: type `A' and `B'. We use B-type masks at all locations apart from the input, where use an A-type mask. We describe these masks in more detail below, and visualise them for a $3 \times 3 \times 3$ kernel in Figure \ref{fig:mask}.
\end{enumerate}

For a three-dimensional kernel of depth $d$, height $h$ and width $w$, consider the following masks that we apply as a point-wise multiplication to the kernel.

\paragraph{A-Type Mask}

\begin{equation}
   M_{d, h, w}  = \begin{cases} 
      1 & \text{if } d \leq \floor{k/2}  \\
      0 & \text{else}
   \end{cases}. \label{eq:conv_op}
\end{equation}

\paragraph{B-Type Mask}

\begin{equation}
   M_{d, h, w}  = \begin{cases} 
      1 & \text{if } d \leq \ceil{k/2}  \\
      0 & \text{else}
   \end{cases}. \label{eq:conv_op}
\end{equation}

\section*{A.3 Visualisation and Coding Scheme for  SHVC-ArIB}

\begin{figure*}[h!]
\centering
\includegraphics[width=.8\textwidth]{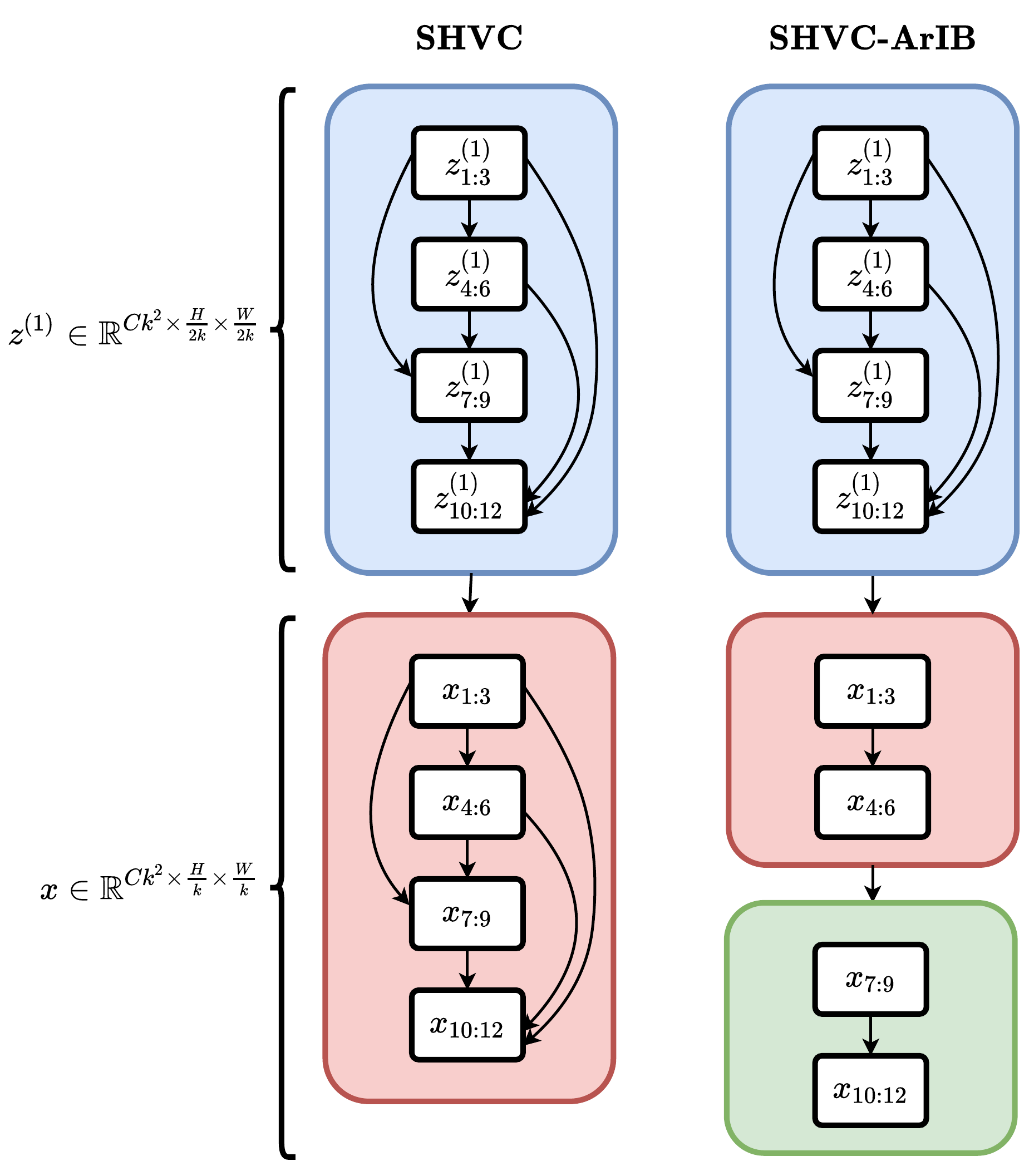}
\caption{A comparison of the factorisation used in the priors of SHVC (left) and SHVC-ArIB (right). Variable groupings are represented by coloured blocks. Arrows indicate explicit dependencies. In SHVC-ArIB, there is no direct link between $z^{(1)}$ (blue) and $x_{7:12}$ (green).}
\label{fig:prior}
\end{figure*}

\begin{figure*}[h!]
\centering
\includegraphics[width=.8\textwidth]{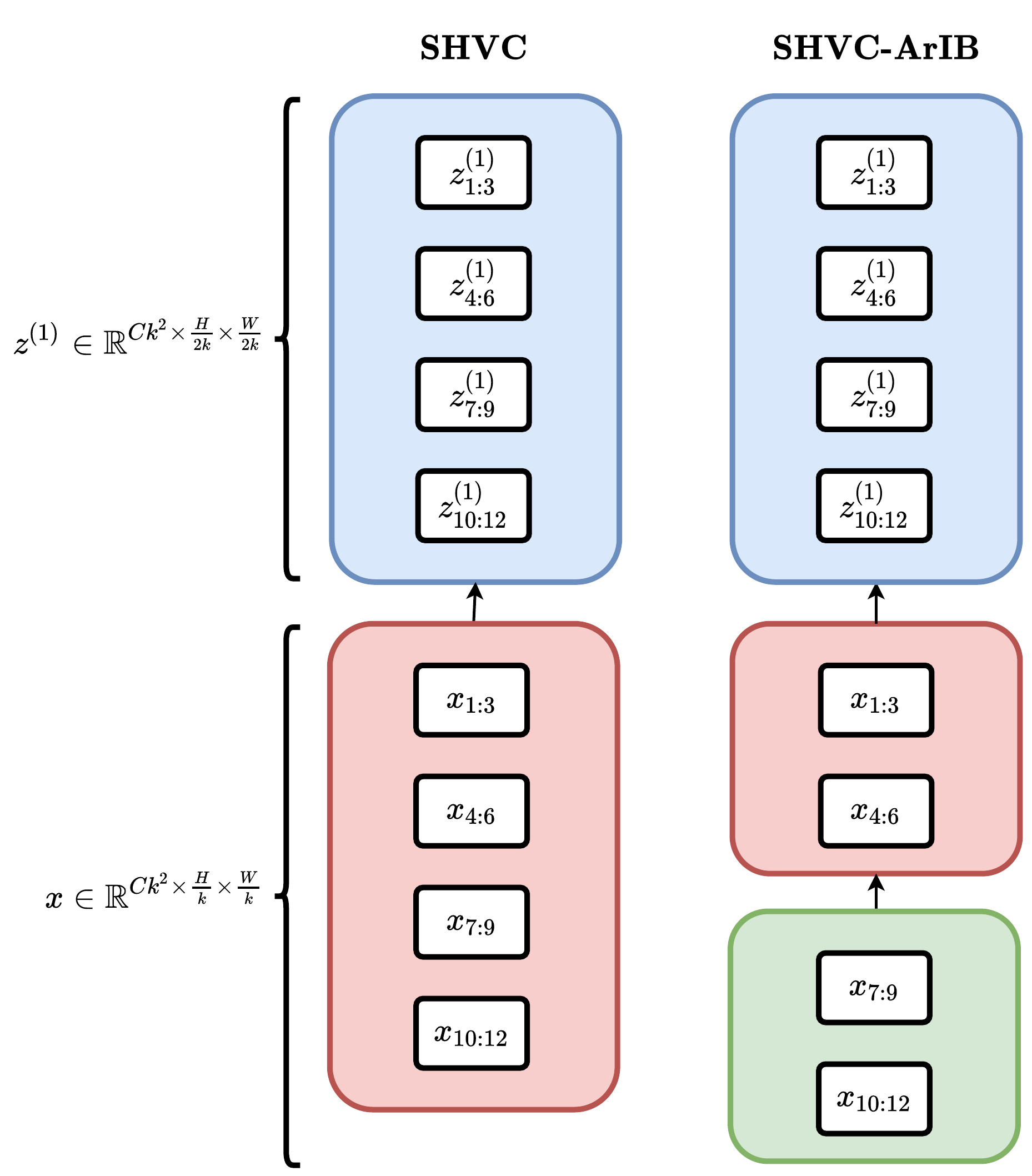}
\caption{A comparison of the factorisation used in the posteriors of SHVC (left) and SHVC-ArIB (right). Variable groupings are represented by coloured blocks. Arrows indicate explicit dependencies. In SHVC-ArIB, there is no direct link between $z^{(1)}$ (blue) and $x_{7:12}$ (green).}
\label{fig:posterior}
\end{figure*}

\paragraph{Variable Dependencies} In Figure \ref{fig:prior} and Figure \ref{fig:posterior}, we illustrate the differences in the dependency structures in the priors and posteriors of SHVC and SHVC-ArIB. 
For ease of presentation, we do so using one latent variable (i.e. $L=1$) and assume $k=2$. 
We further assume $z^{(1)}$ is two-times smaller in spatial resolution that $x$ but has the same number of channels, i.e. $C=3$.

\paragraph{Coding Schemes} Here we use the above model specification as an example to illustrate the encoding and decoding processes of SHVC and SHVC-ArIB. Encoding and decoding algorithms for SHVC can be found in Algorithms \ref{alg:enc} and \ref{alg:dec}. Encoding and decoding algorithms for SHVC-ArIB can be found in Algorithms \ref{alg:arib_enc} and \ref{alg:arib_dec}.  
At the global level, the coding algorithm is consistent with that of Bit-Swap, and at the local level, the encoding of slices in latent and the data is conducted in the reverse order.
Since the above model only involves one latent variable, the global level Bit-Swap degenerates to the original bb-ANS.

\begin{algorithm}
\caption{SHVC Encoding}\label{alg:shvc-encoding}
\begin{algorithmic}
\STATE {\textbf{Input:} data to compress $x$}
\STATE {\textbf{Step 0:} Get auxiliary initial bits $c_0$}
\STATE {\textbf{Step 1:} Decode $z^{(1)}$ with $q(z^{(1)}|x)$}
\STATE {\textbf{Step 2:} Encode $x$ with $p(x|z^{(1)})$}
\STATE {\quad Encode $x_{12}$ with $p(x_{12}|x_{1:11}, z^{(1)})$}
\STATE {\quad Encode $x_{11}$ with $p(x_{11}|x_{1:10}, z^{(1)})$}
\STATE {\quad Encode $x_{10}$ with $p(x_{10}|x_{1:9}, z^{(1)})$}
\STATE {\quad ...}
\STATE {\quad Encode $x_{2}$ with $p(x_{2}|x_{1}, z^{(1)})$}
\STATE {\quad Encode $x_{1}$ with $p(x_{1}|z^{(1)})$}
\STATE {\textbf{Step 3:} Encode $z^{(1)}$ with $p(z^{(1)})$}
\STATE {\quad Encode $z^{(1)}_{12}$ with $p(z^{(1)}_{12}|z^{(1)}_{1:11})$}
\STATE {\quad Encode $z^{(1)}_{11}$ with $p(z^{(1)}_{11}|z^{(1)}_{1:10})$}
\STATE {\quad Encode $z^{(1)}_{10}$ with $p(z^{(1)}_{10}|z^{(1)}_{1:9})$}
\STATE {\quad ...}
\STATE {\quad Encode $z^{(1)}_{2}$ with $p(z^{(1)}_{2}|z^{(1)}_{1})$}
\STATE {\quad Encode $z^{(1)}_{1}$ with $p(z^{(1)}_1)$}
\STATE {\textbf{Output:} final bit stream $c$}
\end{algorithmic}
\label{alg:enc}
\end{algorithm}

\begin{algorithm}
\caption{SHVC Decoding}\label{alg:shvc-decoding}
\begin{algorithmic}
\STATE {\textbf{Input:} bit stream $c$}
\STATE {\textbf{Step 1:} Decode $z^{(1)}$ with $p(z^{(1)})$}
\STATE {\quad Decode $z^{(1)}_{1}$ with $p(z^{(1)}_1)$}
\STATE {\quad Decode $z^{(1)}_{2}$ with $p(z^{(1)}_{2}|z^{(1)}_{1})$}
\STATE {\quad ...}
\STATE {\quad Decode $z^{(1)}_{10}$ with $p(z^{(1)}_{10}|z^{(1)}_{1:9})$}
\STATE {\quad Decode $z^{(1)}_{11}$ with $p(z^{(1)}_{11}|z^{(1)}_{1:10})$}
\STATE {\quad Decode $z^{(1)}_{12}$ with $p(z^{(1)}_{12}|z^{(1)}_{1:11})$}
\STATE {\textbf{Step 2:} Decode $x$ with $p(x|z^{(1)})$}
\STATE {\quad Decode $x_{1}$ with $p(x_{1}|z^{(1)})$}
\STATE {\quad Decode $x_{2}$ with $p(x_{2}|x_{1}, z^{(1)})$}
\STATE {\quad ...}
\STATE {\quad Decode $x_{10}$ with $p(x_{10}|x_{1:9}, z^{(1)})$}
\STATE {\quad Decode $x_{11}$ with $p(x_{11}|x_{1:10}, z^{(1)})$}
\STATE {\quad Decode $x_{12}$ with $p(x_{12}|x_{1:11}, z^{(1)})$}
\STATE {\textbf{Step 3:} Encode $z^{(1)}$ with $q(z^{(1)}|x)$}
\STATE {\textbf{Output:} auxiliary initial bit stream $c_0$, data to decompress $x$}
\end{algorithmic}
\label{alg:dec}

\end{algorithm}

\begin{algorithm}
\caption{SHVC-ArIB Encoding}\label{alg:shvc-arib-encoding}
\begin{algorithmic}
\STATE {\textbf{Input:} data to compress $x$}
\STATE {\textbf{Step 0:} Get autoregressive initial bits by encoding $x_{7:12}$}
\STATE {\quad Encode $x_{12}$ with $p(x_{12}|x_{1:11})$}
\STATE {\quad ...}
\STATE {\quad Encode $x_{7}$ with $p(x_{7}|x_{1:6})$}
\STATE {\textbf{Step 1:} Decode $z^{(1)}$ with $q(z^{(1)}|x_{1:6})$}
\STATE {\textbf{Step 2:} Encode $x_{1:6}$ with $p(x_{1:6}|z^{(1)})$}
\STATE {\quad Encode $x_{6}$ with $p(x_{6}|x_{1:5}, z^{(1)})$}
\STATE {\quad ...}
\STATE {\quad Encode $x_{1}$ with $p(x_{1}|z^{(1)})$}
\STATE {\textbf{Step 3:} Encode $z^{(1)}$ with $p(z^{(1)})$}
\STATE {\quad Encode $z^{(1)}_{12}$ with $p(z^{(1)}_{12}|z^{(1)}_{1:11})$}
\STATE {\quad Encode $z^{(1)}_{11}$ with $p(z^{(1)}_{11}|z^{(1)}_{1:10})$}
\STATE {\quad Encode $z^{(1)}_{10}$ with $p(z^{(1)}_{10}|z^{(1)}_{1:9})$}
\STATE {\quad ...}
\STATE {\quad Encode $z^{(1)}_{2}$ with $p(z^{(1)}_{2}|z^{(1)}_{1})$}
\STATE {\quad Encode $z^{(1)}_{1}$ with $p(z^{(1)}_1)$}
\STATE {\textbf{Output:} final bit stream $c$}
\end{algorithmic}
\label{alg:arib_enc}
\end{algorithm}

\begin{algorithm}
\caption{SHVC-ArIB Decoding}\label{alg:shvc-arib-decoding}
\begin{algorithmic}
\STATE {\textbf{Input:} bit stream $c$}
\STATE {\textbf{Step 1:} Decode $z^{(1)}$ with $p(z^{(1)})$}
\STATE {\quad Decode $z^{(1)}_{1}$ with $p(z^{(1)}_1)$}
\STATE {\quad Decode $z^{(1)}_{2}$ with $p(z^{(1)}_{2}|z^{(1)}_{1})$}
\STATE {\quad ...}
\STATE {\quad Decode $z^{(1)}_{10}$ with $p(z^{(1)}_{10}|z^{(1)}_{1:9})$}
\STATE {\quad Decode $z^{(1)}_{11}$ with $p(z^{(1)}_{11}|z^{(1)}_{1:10})$}
\STATE {\quad Decode $z^{(1)}_{12}$ with $p(z^{(1)}_{12}|z^{(1)}_{1:11})$}
\STATE {\textbf{Step 2:} Decode $x_{1:6}$ with $p(x_{1:6}|z^{(1)})$}
\STATE {\quad Decode $x_{1}$ with $p(x_{1}|z^{(1)})$}
\STATE {\quad ...}
\STATE {\quad Decode $x_{6}$ with $p(x_{6}|x_{1:5}, z^{(1)})$}
\STATE {\textbf{Step 3:} Encode $z^{(1)}$ with $q(z^{(1)}|x_{1:6})$}
\STATE {\textbf{Step 4:} Decode $x_{7:12}$ with $p(x_{7:12}|x_{1:6})$}
\STATE {\quad Decode $x_{7}$ with $p(x_{7}|x_{1:6})$}
\STATE {\quad ...}
\STATE {\quad Decode $x_{12}$ with $p(x_{12}|x_{1:11})$}
\STATE {\textbf{Output:} data to decompress $x$}
\end{algorithmic}
\label{alg:arib_dec}
\end{algorithm}

\section*{A.4 SHVC Architecture and Experimental Details}

For both our encoder and decoder architectures, we use 8-layer Residual networks with PReLU 
activation units and weight-normalization. For CIFAR10 we additionally use dropout layers between residual connections to prevent overfitting.

To highlight the effectiveness of our approach, we additionally train a small "Lite" model, which uses four fully-connected convolutional layers. Here we reduce the number of channels as we downsample the latent variables across layers. These are detailed as follows:

\begin{itemize}
    \item $p(x|z^{(1)})$ uses 32 channels.
    \item $p(z^{(1)}|z^{(2)})$ uses 24 channels.
    \item $p(z^{(2)}|z^{(3)})$ uses 16 channels.
    \item $p(z^{(3)}|z^{(4)})$ uses 8 channels, if it exists. 
\end{itemize}

For training, we use the Adam optimizer with default learning settings and an initial learning rate of $5\times 10^{-4}$. We exponentially anneal this learning rate to $1 \times 10^{-5}$ during training. We further use gradient-clipping to control for numerical stability. 

We run all of our experiments on a single NVIDIA Tesla V100. 

\end{document}